\documentclass{aastex}          

\begin{document}
%
\title{The formation of planetary disks and winds: an ultraviolet view}

\shorttitle{UV view of young solar systems}
\shortauthors{Gomez de Castro}

\author{Ana I. Gomez de Castro\altaffilmark{1}}

\altaffiltext{1}{Fac. de CC Matematicas, Universidad Complutense de Madrid,
28040 Madrid, Spain}

\email{aig@mat.ucm.es}

\begin{abstract}

Planetary systems are angular momentum reservoirs generated during
star formation. This accretion process produces very powerful
engines able to drive the optical jets and the molecular outflows.
A fraction of the engine energy is released into heating thus the
temperature of the engine ranges from the 3000K of the inner disk
material to the 10MK in the areas where magnetic reconnection
occurs. There are important unsolved problems concerning the
nature of the engine, its evolution and the impact of the engine
in the chemical evolution of the inner disk. Of special relevance
is the understanding of the shear layer between the stellar
photosphere and the disk; this layer controls a significant
fraction of the magnetic field building up and the subsequent
dissipative processes ougth to be studied in the UV.

This contribution focus on describing the connections between
1 Myr old suns and the Sun and the requirements for new
UV instrumentation to address their evolution during this period.
Two types of observations are shown to be needed: monitoring
programmes and high resolution imaging down to, at least,  milliarsecond
scales.

\end{abstract}

\keywords{}

\section{Introduction}

Solar sytem progenitors, from the deeply embedded sources to
the weak line T Tauri Stars (WTTSs), are sources of high
energy radiation (X-ray to UV); the total energy
radiated in this range goes from $\sim 0.02$~L$_{\odot}$
measured in the very young
sources through their X-ray radiation to the 0.2 L$_{\odot}$ radiated in the
UV during the T Tauri phase (or Phase~TT) (Preibish 2004, G\'omez de Castro 2008).
Later on, in the Weak line T Tauri Phase the energy released drop to
$\sim 10^{-3}$~L$_{\odot}$ radiated in X-ray. Being TTSs
intrinsically cool stars (with $\log T_{\rm eff}\sim$ 6500-3600~K),
surrounded by cool accretion disks radiating at infrared
wavelengths, the source
of this energy must be searched in the release of magnetic energy.

It is well known that the mediation of magnetic fields in the
accretion process is able to heat up the plasmas since a fraction of the
gravitational energy lost during accretion is invested in field
amplification and dynamo action thus, radiative loses are pushed
towards the high energy range. Unfortunately, in the early phases
(ages $< 0.1$Myr) most of this radiation is reabsorbed by the
dense circumstellar environment (Av$>3$) and only the hard X-ray
radiation is able to escape from the system providing direct
information on the evolution of the accretion process. After 1
Myr, extinction drops enough to make the engine accessible to UV
wavelengths; current technologies allow to carry out high
resolution spectroscopy in the UV range which is also extremely
rich in spectral tracers so a single echelle spectra can provide
information on molecular, atomic and ionized gas (from singly
ionized gas to very high ionization species such as  Fe~XII,
Fe~XVIII or  Fe~XXI). Thus, UV spectroscopy is an extremely
efficient tool to study solar system progenitors from 1 Myr on with the
current technology and these application will be discussed in detail
below.  However, one can foresee a future when
microarcseconds UV imaging  will be available and
studies similar to those being run on the Sun, will be feasible
from 1 Myr old Suns all the way down into the main sequence while
the young planetary disk settles down and life begins to grow.
This review deals with this; with a description of our
current understanding of the evolution from the T Tauri phase to
the modern Sun and with a non-technologically biased ambitious
view of what we could learn from challenging new UV observatories.
This review has been written after the end of the 1st. conference
of the network for UV astronomy held in El Escorial in May 2007
where some challenging projects for new space observatories were
presented; you should find references to some of them in this text.

\section{Physics to be understood I: the gravito-magnetic engine}

During the phase TT, stars count on an energy source
which is not available during the main sequence evolution: gravitational
energy from the infalling material. This extra energy is released either
through shocks on the stellar surface or through the gravito-magnetic
interaction between the star and the disk.

Shocks release the kinetic energy of the infalling material into
heating at the impact point. If matter infall occurs along the
field lines, all the gravitational energy is damped into heating
and the gas may reach temperatures as high as $\sim 10^6$K. The
dominant output radiation is produced by the photoionized preshock
infalling gas radiating mainly in the UV range (G\'omez de Castro
\& Lamzin 1999; Gullbring et al. 2000). As the density of the
infalling gas column is high ($n_e \simeq 10^9-10^{12}$~cm$^{-3}$)
the thickness of the radiating column is expected to be negligible
compared with the stellar radius thus, accretion shocks are
observed as {\it hot spots} on the stellar surface. As such, they
are expected to produce a rotationally modulated signal that has
been detected in monitoring campaigns of some stars both in optical 
(see i.e. Petrov et al. 2001;
Bouvier et al. 2003) and UV (G\'omez de Castro \& Fern\'andez
1996; G\'omez de Castro \& Franqueira 1997a). An important result 
of these campaigns is that only $\sim
50$\% of the UV continuum excess is rotationally modulated. Thus,
a significant fraction of the UV excess is not produced by the
accretion shocks even in sources where rotational modulation
has been detected. However, this excess decreases as the star
approaches the main sequence (see Fig.~1).

\begin{figure}[]
\centerline{\includegraphics[width=18pc]{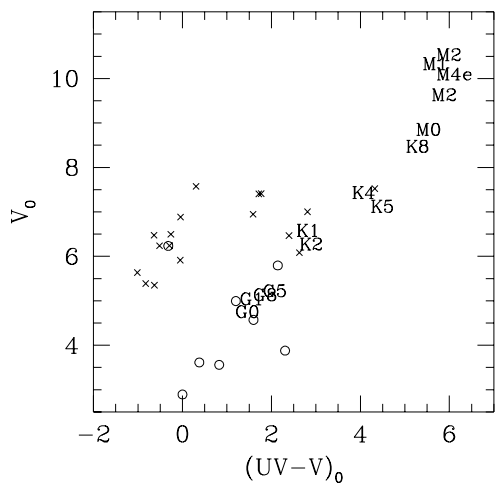}}
\caption{The (UV-V, V) colour -- magnitude diagram for the
T Tauri stars observed with the IUE satellite in the Taurus region.
The crosses represent cool TTSs (spectral types later than
$\sim $ K3) and the open circles warm TTSs (spectral types
earlier than $\sim$ K3). The location of the main sequence is
marked by the spectral types. The stars closer to the main sequence
are the WTTSs (from G\'omez de Castro 1997).}
\end{figure}

In fact, the major source of high energy radiation is the dissipation
of the magnetic and mechanical energy produced by the gravito-magnetic
engine. A simple analogy can be made with a self-regulated hydraulic
turbine: the potential energy of the gas falling from the disk into the
stellar magnetic field drives to the generation of electric currents
due to the Lorentz force that, in turn, create new field components
through dynamo action. There are however, a great number of uncertainties
in the way the system self-regulates and also on the dependence of the
engine details on initial conditions such as the effective gravity of
the star, the role of stellar radiation and magnetic field on the engine
performance and the role of the ionizing radiation produced by the engine on
the evolution of the mass storage, the disk.

The Sun, itself, provides important clues to understand the
physics of the gravito-magnetic engine and its evolution. At the
base of the Sun convective layer, the tachocline marks the
location of the shear layer between the rigid body rotation of the
radiative core and the differentially rotating convective
envelope. The tachocline is significantly prolate; it is centered
at 0.69R$_{\odot}$ at the equator and 0.72R$_{\odot}$ at latitude
60$^o$ (Basu \& Antia, 2003). The tachocline thickness is $\sim
0.04$R$_{\odot}$. The angular velocity profile based on
helioseismic inversions shows that at a latitude of about 35$^o$,
the radial gradient changes sign becoming negative for latitudes
$> 35^o$. This latitude marks the limit of the two latitude belts
where the overwhelming majority of sunspots occur. There are also
some indications of the meridional flow moving equatorwards below
this latitude and polewards above it (see Miesch 2005). The Solar
wind is (magnetic)
latitude dependent during solar minimum;
above $\sim 35^o$ is fast (1000~km/s) and
thin, below is slower (300~km/s) and denser (see Fig.2 from
Ulysses data).  The current paradigm
for how solar dynamo operates includes: (1) field amplication in a
turbulent downflow ($\alpha$ effect) that is pumped downward by
convection and accumulate in the overshoot region and the
tachocline; (2) field amplification and organization into toroidal
flux tubes and sheets by differential rotation in the tachocline;
(3) magnetic instabilities (buoyancy) drives the field to the surface and
(4) the Coriolis force acting on the rising structures depends on the
latitude producing a latitude-dependent emergence of bipolar magnetic
structures.

\begin{figure}[]
\centerline{\includegraphics[width=18pc]{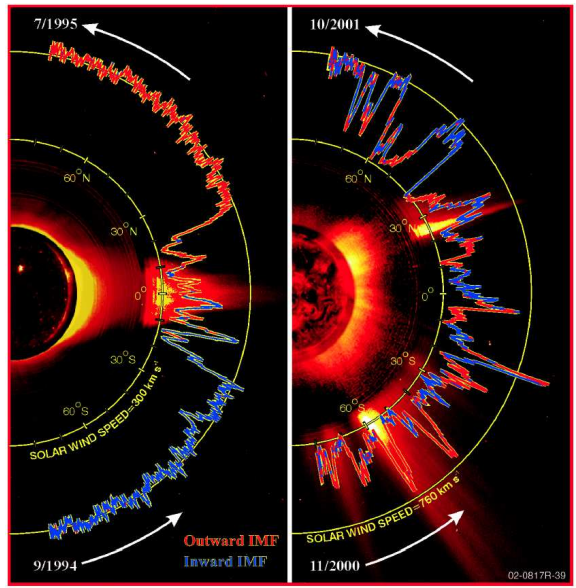}}
\caption{{\it ``Solar wind observations collected by the Ulysses spacecraft
during two separate polar orbits of the Sun, six years apart,
at nearly opposite times in the solar cycle. Near solar minimum
(left) activity is focused at low altitudes, high-speed solar wind
prevails, and magnetic fields are dipolar. Near solar maximum (right),
the solar winds are slower and more chaotic, with fluctuating magnetic
fields.''} (From NASA Solar Probe Web (solarprobe.gsfc.nasa.gov),
courtesy of Southwest Research Institute and the Ulysses/SWOOPS team)
}
\end{figure}

Going backwards in time, during the phase TT,
the problem becomes complicated by the
presence an additional ``tachocline'' or differentially rotating
region attached to the convective layer. This {\it external
tachocline} connects the star with the accretion disk which
rotates significantly faster than the stellar surface; rotation
periods during the TT phase are about 7-8 days ($\Omega _* =
0.8-0.9$ day$^{-1}$) while the Keplerian frequency is:
$$
\Omega _k = 11.1  {\rm day}^{-1} \lgroup \frac {M}{M_{odot}} \rgroup ^{1/2}
\lgroup \frac {r}{3 R_{odot}} \rgroup ^{-3}
$$
Keplerian disk corotation radius is at,
$$
r_{\rm co} = (7.2 - 6.9)R_{\odot}  \lgroup \frac {M}{M_{odot}} \rgroup ^{1/2}
$$
To avoid this large shear, the magnetosphere will grow to balance the
toroidal component of the flux with the angular momentum of the infalling
matter (Ghosh \& Lamb 1979) thus,
$$
\frac{B_p B_t}{4 \pi} 4\pi r^2 \Delta r \simeq \dot M r V_k
$$
where $B_p$ and $B_t$ are the poloidal and toroidal components of the
field respectively, $r$ is the magnetosphere radius, $\Delta r$ is
the thickness of the shear layer, $\dot M$ is the accretion rate and
$V_k$ is the Keplerian velocity at the magnosphere radius. For typical
T~Tauri stars parameters:
$$
r_{\rm mag}= 4.4 R_{\odot} \gamma ^{2/7} \lgroup \frac {B_*}{1kG}
\rgroup^{4/7} \lgroup \frac {\dot M}{10^{-8} M_{\odot} {\rm
yr}^{-1}} \rgroup^{-2/7} \lgroup \frac {M_*}{M_{\odot}}
\rgroup^{-1/7}
$$
where $\gamma ^{2/7}$ is a factor about unity ($\gamma =
(B_t/B_p)(\Delta r /r)$, see Lamb, 1989). Notice that the main
uncertainties in the physics, namely the ratio between the
toroidal and the poloidal components and the relative thickness of
the {\it ``external tachocline''} are enclosed in this factor.
As in the Solar
interior, the shear region is fed by turbulent, magnetized
material though this comes from the accretion disk instead of
the convective layer. The turbulent disk dynamo is fed by  the
magneto-rotational instability in the acretion disk. Shear
amplifies the field producing a strong toroidal component; an
external dynamo sets in. This toroidal field and
the associated magnetic pressure push the field lines outwards
from the disk rotation axis, inflating and opening  them in a {\it
butterfly-like pattern} reminiscent of the helmet streamers in
the solar corona, so producing a current layer between the
stellar and the disk dominated regions as displayed in Fig~3.
Magnetic field dissipation in the current layer produces high
energy radiation and particles. The magnetic link between the star
and the disk is broken and reestablished continuously by magnetic
reconnection. The opening angle of the current layer, as well as
its extent, depends on the stellar and disk fields, the accretion
rate and the ratio between the inner disk radius and the stellar
rotation frequencies. Hot, pressure driven outflows are produced
from the region closer to the rotation axis while cool
centrifugally driven flows are produced by the disk; plasmoids are
ejected from the current layer generating a third outflowing
component.

\begin{figure}[]
\centerline{\includegraphics[width=18pc]{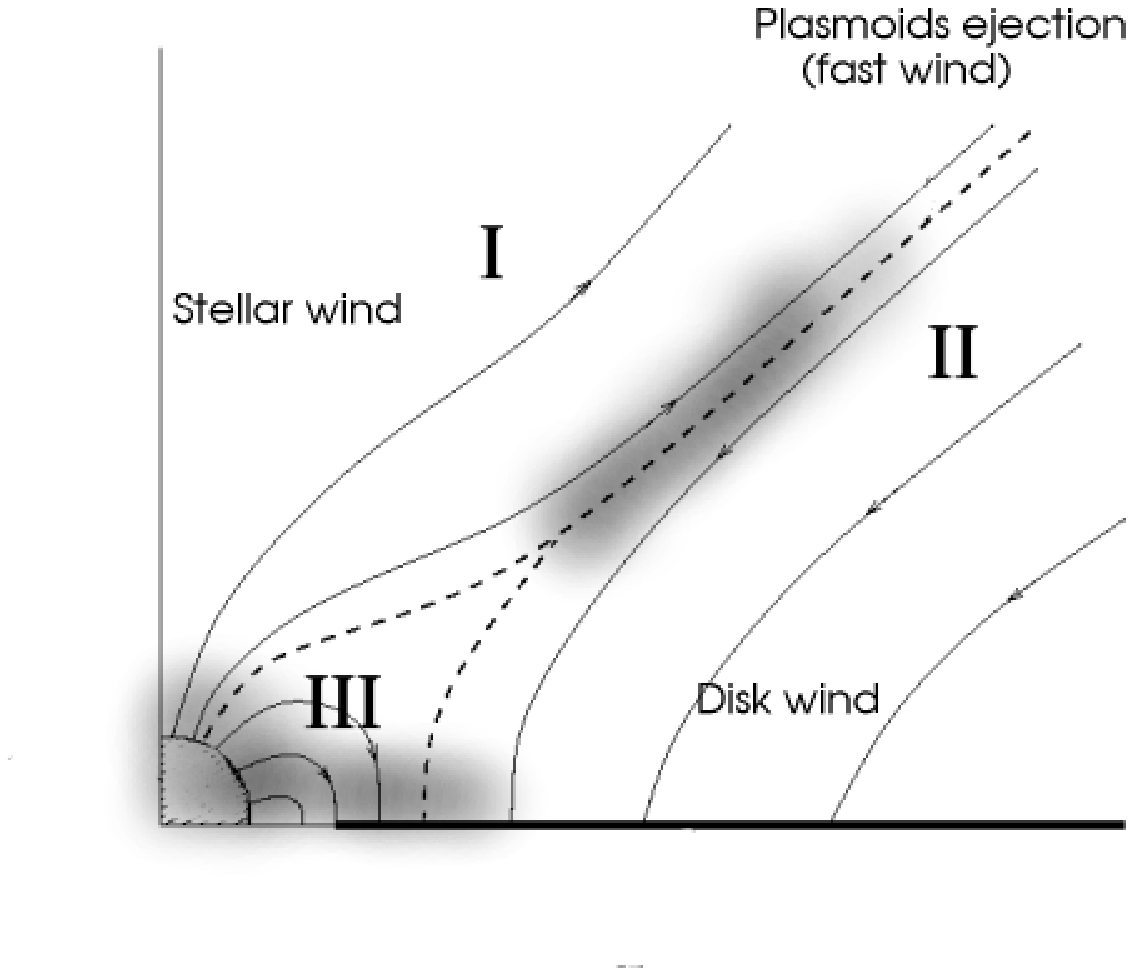}}
\caption{The interaction between the stellar magnetic field and the
disk twists the stellar field lines due to the differential rotation.
The toroidal magnetic field generated out of the poloidal flux and
the associated pressure tends to push the field lines outwards,
inflating them, and eventually braking the magnetic link between
the star and the disk (boundary between regions I and II).
Three basic regions can be defined:
Region I dominated by the stellar wind, Region II dominated by
the disk wind and Region III dominated by stellar magnetospheric
phenomena. The dashed line traces the boundaries between this
three regions. The continuous lines indicate the topology
of the field and the shadowed areas represent regions where
magnetic reconnection events are likely to occur, producing
high energy radiation and particles (from G\'omez de Castro 2004).}
\end{figure}

Disk-star interaction has been investigated by  means of numerical
simulations since the early works by Goodson et al (1997) till the
last results ({\it i.e.} von Rekowski \& Brandenburg 2006). They
show that the fundamental  mechanism for disk winds formation is
robust; numerical simulations with different parameters (disk/star
fields) and initial conditions produce disk winds. Stellar winds
are much more sensitive to the physical conditions and specially
to the stellar field; compare the results of simulations assuming
that the stellar field is a magnetic dipole (von Rekowski \&
Brandenburg 2004) with those of simulations where the stellar
field is prescribed through the action of the stellar dynamo (von
Rekowski \& Brandenburg, 2006). In fact, the characteristics of
the accretion flow  and the winds (dominant driver, temperature,
terminal velocity, density, variability) depend on the physical
properties of the system such as the degree of magnetization of
the disk, the characteristics of the disk dynamo and  the stellar
field.

The bulk of the energy produced in this engine is released at UV
and X-ray wavelenghts as in the Sun atmosphere. In the very early
epochs, when extinction is high ($A_V \geq 3$), only the X-ray
radiation from the engine is detected. Later on, about 1 Myr,
extinction drops and the engine can be studied in the UV. Only in
the UV, the various components of the engine can be defined and
studied as well as their evolution, starting during  the phase~TT
all the way down into the main sequence.

\section{Physics to be understood II: the
impact of the engine on disk evolution}

Though often neglected, the impact of the engine in the inner disk
evolution is enormous. On the one hand, the engine adds a significant
poloidal component in the inner disk thus favouring gas motions
perpendicular to the disk as shown by the numerical simulations,
on the other hand, the engine is a source of highly energetic
radiation where part of the dissipation is produced at heights of
some few stellar radii above the disk in the inflating current
layer; this high latitude illumination favours energy absorption
by the disk. Both together act to increase the disk scale height and
the absorption of the radiation henceforth producing the
rarification of the disk
atmosphere and favoring disk evaporation close to the star.
To achieve the evaporation of a standard optically thin accretion disk,
the sound speed should be comparable to the keplerian velocity thus,
$$
T = \frac{G M_*/r_{\rm mag}}{\mu m_H / \gamma \kappa}
= 3.13 \times 10^7 K \lgroup \frac {M_*}{M_{\odot}}
\rgroup \lgroup \frac {r_{\rm mag}}{4.4 R_{\odot}}
\rgroup ^{-1}
$$
where $\mu$ is the mean molecular weight, $\gamma $ is the
polytropic index and $\kappa $ is Boltzmann constant.
However, this value relaxes in the presence of a poloidal
field as the expected in the disk-magnetosphere interface so,
$$
T = 3.13 \times 10^7 K \frac {\beta}{1+\beta}
\lgroup \frac {M_*}{M_{\odot}}\rgroup
\lgroup \frac {R_{\rm mag}}{4.4 R_{\odot}}
\rgroup ^{-1}
$$
where $\beta$ is the rate between magnetic and thermal pressures.
Thus, for highly magnetized environments, $T$ may drop
arbitrarily. For thin accretion disks\footnote{See Frank et al 2002
for the standard prescription of the thin disk density and temperature
as a function the accretion rate, radius and stellar mass.} penetrated
by the stellar dipolar field, $B_*$,

\begin{eqnarray}
\beta &=&  \frac {\gamma \kappa T/ \mu M_H}{B^2 /4 \pi \rho} \\
&=&
4.77  \lgroup \frac {M_*}{M_{\odot}}\rgroup ^{7/8}
\lgroup \frac {\dot M}{10^{-8}M_{\odot}{\rm yr}^{-1}}\rgroup ^{17/20} \\
&\times & \lgroup \frac {r}{4.4 R_{\odot}}\rgroup ^{-21/8}
\lgroup \frac {B_*}{1kG}\rgroup ^{-2} \\
\end{eqnarray}
\noindent
where $r$ is the disk radius, $B_*$ the stellar magnetic field and
$\dot M$ the accretion rate. Note that
$\beta $ drops to 0.02 for accretion rates of $10^{-9}$M$_{\odot}$yr$^{-1}$.

 Another important phenomenon to be considered is that the disk
is not unlocked from the engine so disk material should be
subjected to the propagation of the Alfv\'en waves, shear waves
and global alfv\'en oscillations driven from the interface.
In summary, we might expect the inner rim of the disk to be hot
with temperatures of about $10^4$K well above the
temperature of dust sublimation.

The role of far-UV radiation fields and high energy particles in
the disk chemical equilibrium is now beginning to be understood.
Bergin et al. (2003) showed how strong Ly$\alpha$ emission may
contribute to the observed enhancement of CN/HCN in the disk. The
penetration of UV photons coming from the engine in a dusty disk
could produce an important change in the chemical composition of
the gas allowing the growth of large organic molecules. In this
context, UV photons photodissociating organic molecules at
$\lambda > 1500$~\AA\ could play a key role in the chemistry of
the inner regions of the disk, while those photodissociating H$_2$
and CO will control the chemistry of the external layers of the
disk directly exposed to the radiation from the central engine.

\section{Lessons learned from UV (spectroscopic) observations}

The first observations of pre-main sequence (PMS) stars in the UV
were carried out with the International Ultraviolet Explorer (IUE)
(1979-1997). The observations showed that pre-main sequence stars
have UV fluxes exceeding those of main sequence stars by a factor
of about 50. In fact, the UV excess decreases as the stars
approach the main sequence as shown in Fig~1.

UV radiation provides direct information on the interaction
between the star and the disk. This includes all the various
components mentioned above: the shear layer, i.e.the {\it external
tachocline}, the
wind, the enhanced magnetospheres, mass ejections from
reconnecting loops, shocks between the various wind components
(among themselves and also with the disk material) and as well as
the inner regions of the disk. There is a recent review on UV
observations of pre-main sequence stars and young planetary disks
(G\'omez de Castro et al 2006) where a detailed accounting of the
work carried out since the IUE times is summarized. Thus, I should
concentrate on the main lessons learned from IUE and HST
observations\footnote{Some observations have also been obtained
with FUSE but its small effective area has allowed to observed
only the brightest of the TTSs and some Vega-like disks.} that are:

\subsection{About the accretion flow}

The actual measurements of infalling gas in the UV are scarce. The
are hints of accretion in the large extent of the red wings of the
main UV resonance lines (Herczeg et al 2005) or through the
detection of redshifted absorption components on the profiles
of the most prominent emission lines.
However, the only target for which there is clear spectroscopic
evidence of accretion shocks (in the UV) is RY~Tau (see Fig.~4).
Two observations of the same star obtained in 1993 and 2001 show that
there is a variable redshifted component. From this single
observation three important properties are learnt:
\begin{enumerate}
\item As described above, UV radiation from accretion shocks
is predicted to be produced in the preshock region on scale
heights significantly smaller than the stellar radius. Thus, it is
expected that only matter falling onto the visible hemisphere can
be detected at UV wavelengths. The fact that the variable flux
component is redwards shifted supports these theoretical
expectations. Moreover, the broadening shows that infalling matter
should cover a significant fraction of the hemisphere to account
for the broad distribution of projected velocities in the
infalling gas.

\item The red wing extends to
velocities of 250~km/s which corresponds to the free-fall velocity\footnote{
RY Tau mass is 1.63~M$_{\odot}$ and radius 2.4~R$_{\odot}$ according to
Hartigan et al 1995} from 1.7 R$_*$ which is much smaller than
the fiducial values derived for the inner disk radius.

\item The UV excess is not only produced by accretion;
also the wind contributes to it. Thus accretion rates derived from
the UV excess, assuming that it is caused just by magnetospheric
infall, are overestimated.

\end{enumerate}

It also adds to our understanding of {\it magnetospheric
accretion}. Magnetospheric accretion was originally proposed to
explain the large broadening of the TTSs H$\alpha $ lines
(Muzerolle et al 1998) later detected also in the UV
resonance lines of CIV, CIII (Ardila et al 2002, Herczeg et al 2005).
Typical line widths are about
200-300 km/s that exceed by far what expected from the rotation
velocities of the TTSs ($10-20$ km/s) even if the corotating
magnetosphere is postulated to extend to some 4-5 stellar radii.
Infall adds a radial velocity component to rotation of the
radiating gas. As free fall velocity is:
$$
v_ {ff} \simeq 315 {\rm km s} ^{-1} \lgroup \frac {M_*}{M_{\odot}}
\rgroup ^{1/2} \lgroup \frac {R_*}{R_{\odot}} \rgroup ^{-1/2}
$$
The observed
profiles broadenings can be reproduced without difficulty. The
observed broadening of H$\alpha$ or Mg~II lines do not vary
significantly requiring that, at least, the spatial average of the
accretion flow is rather stable.  Strong variations in the
accretion flow, like the reported from RY~Tau UV observations, should
also show in the large scale magnetosphere tracers (i.e.
H$\alpha$ or Mg~II profiles).

\begin{figure}[]
\centerline{\includegraphics[width=18pc]{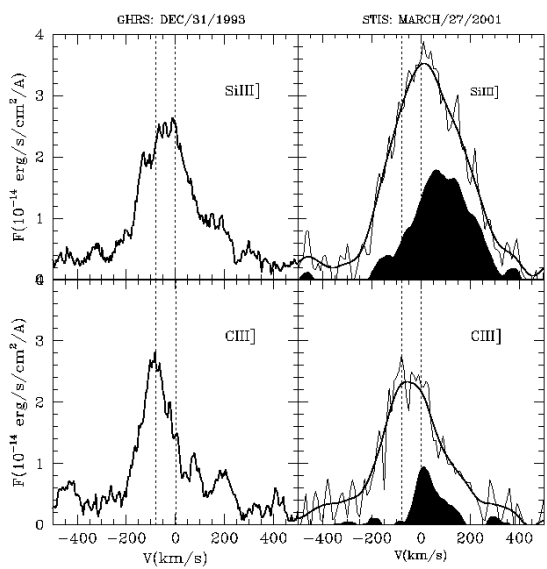}}
\caption{ SiIII] and CIII] UV lines observed in RY~Tau (from
G\'omez de Castro \& Verdugo 2007); RSDP processed data are
plotted with a thin line and the 3-pixels average profile with a
thick line. The rest wavelength of the lines and the velocity of
the unresolved jet at $\simeq -80$ km/s (from G\'omez de Castro \&
Verdugo 2001 and Hamann 1994) are marked with dashed lines. {\it
Left panel}: Observations obtained in Dec. 31st, 1993 with the
GHRS. {\it Right panel}: Observations obtained in March 27th,
2001. Both lines show an excess of flux in the red wing compared
with the 1993 observations; this excess is shaded in the figure.}
\end{figure}

\subsection{About the wind}

There is another possibility to broaden the line profiles: adding
a radial velocity component associated with the outflow. The
presence of magnetic fields and the relevance of centrifugal
launching drives to the formulation of the velocity field in the
wind by means of three components: axial component (along the
rotation axis), radial expansion from the axis and the azimutal
toroidal component (rotation around the axis); in figure 5 there
is a representation of the three components for a warm centrifugal
wind model (from G\'omez de Castro \& Ferro-Font\'an 2005). A
rapid radial expansion close to the star (to guarantee that the
wind density and temperature are about the observed
n$_e=10^{10}$cm$^{-3}$ and T$_e = 20-30 \times 10^{3}$K values)
could produce similar effects in the profiles than those predicted
by magnetospheric infall. Several attempts have been made to
reproduce the wind profiles with cold winds from accretion disks
(Ferro-Font\'an \& G\'omez de Castro 2003), warm disk winds
(G\'omez de Castro \& Ferro-Font\'an, 2005), coronal winds from
accretion disks (Ferreira \& Casse, 2004) and winds driven
by the star-disk interaction (G\'omez de Castro \& von Rekoswki, 2008).

\begin{figure}[]
           \includegraphics[width=18pc]{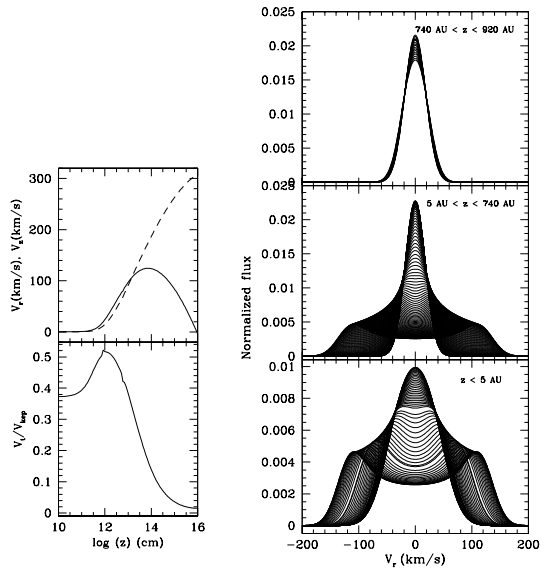}
\caption{The basic kinematics of MHD centrifugal winds is outlined
from the simple semiempirical model of centrifugally driven MHD
winds with thermal pressure of G\'omez de Castro \& Ferro-Font\'an
(2005). {\it Left:} Velocity field in the flow.
$V_r$ (solid) and $V_z$ (dashed) are the components of the
velocity along the $r$
and $z$ axes, respectively. The toroidal component of the
velocity, $V_{t}$ is scaled with respect to the keplerian
velocity of the disk at the radius from which the wind is ejected.
 {\it Right:} Line profiles generated by the wind in ring
of gas  at different heights along the
z-axis. }
\end{figure}

Cold disk winds fail to reproduce the high temperatures observed.
The original wind temperature is as low as the disk one and
heating has to be extracted from photoionization by the central
source. However, this radiation is able to heat the gas only to
mild temperatures of about 10$^4$K. Warm disk winds produce
profiles that are too narrow to reproduce all the observations;
this is because the vertical thermal pressure push forces the
growth of the radial velocity component to heights were the plasma
is already too cool. Finally, winds driven from the star-disk
interaction also produce narrower profiles than the observed in
some sources as shown in Fig.6. According their UV forbidden lines
profiles, TTSs can be classified in two groups: stars with
broadenings with full width half maximum about 150~km~s$^{-1}$
that can be adjusted with the current models and stars with
extremely broad profiles ($> 250$km~s$^{-1}$). The source of the
very large broadenings have to be seek in other structures such as
ion belts or plasma rings as the resolved in RW~Aur (see Sect.
4.3).

\begin{figure}[]
           \includegraphics[width=18pc]{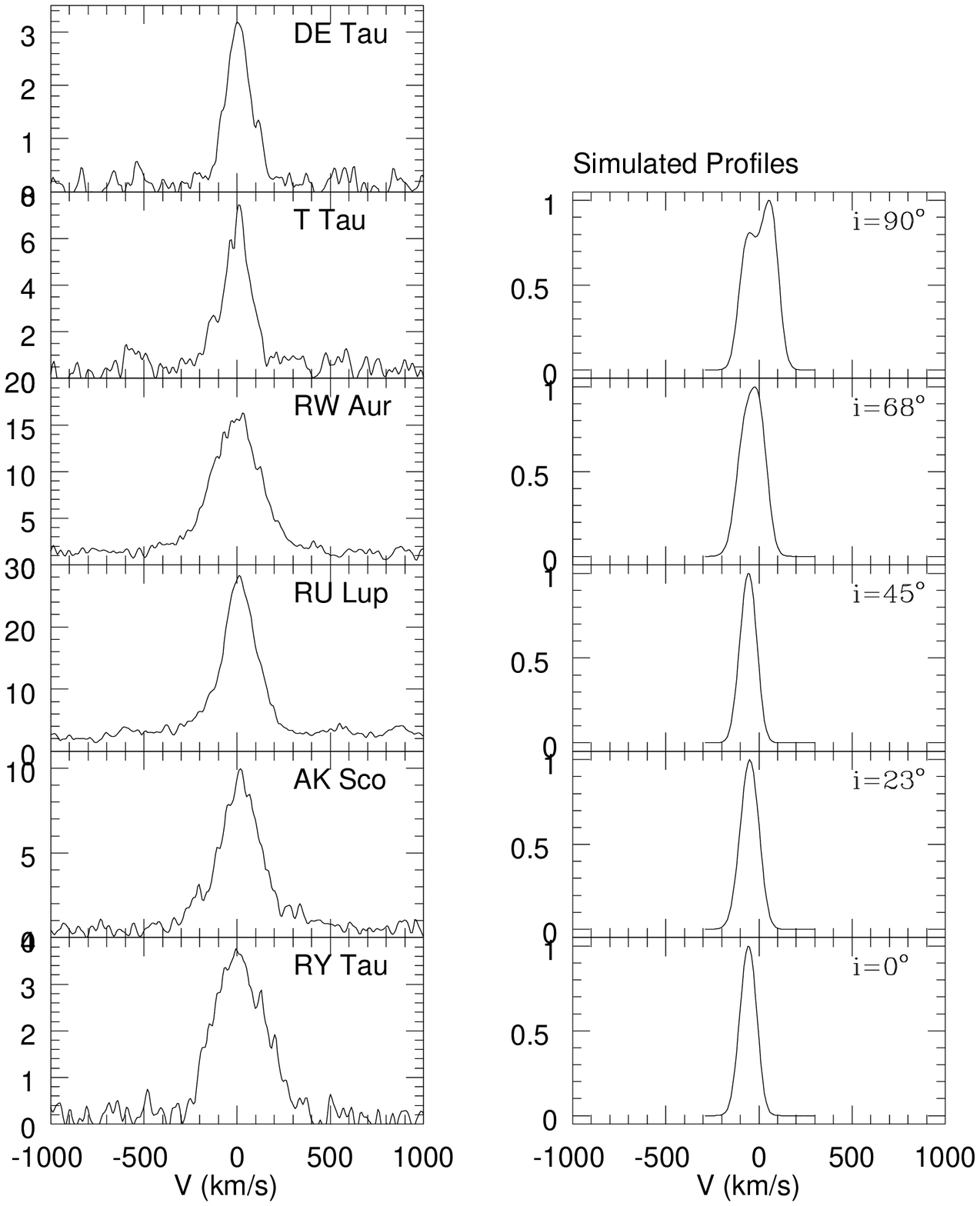}
\caption{{\it Left:} Si~III] profiles of the TTSs observed
with HST; notice the very different line broadenings.
{\it Right:} Predicted Si~III] profiles for winds generated
by means of the interaction between a stellar magnetosphere
with stellar field 1kG and an accretion disk undergoing
$\alpha ^2 \Omega$-dynamo effect (e.g. Krause \& Raedler, 1980)
from G\'omez de Castro \& von Rekowski (2008).
}
\end{figure}

Finally, UV observations have clearly proved that:
\begin{enumerate}
\item  {\it Warm winds are latitude dependent on scales
comparable to the stellar radius}. As an example, the Mg~II
resonance doublet has been observed in a broad sample of 17 TTSs
(adding IUE and HST samples); this is the largest sample of TTSs
observed in a single UV spectral line.  These  lines can be
generically described as broad, asymmetric emission lines with
typical full widths at 10 \% intensity of few hundreds km/s. The
broad blueward shifted absorption component characteristic of
mass-loss was detected in few sources (Penston \& Lago, 1983,
Imhoff \& Appenzeller, 1989) but not in all of them; the degree of
absorption (the asymmetry of the line) varies from not absorption
to full absorption of the bluewards shifted emission (see G\'omez
de Castro, 1997).
\item {\it The collimated flow, the jet, radiates in the UV as
well as the bow-shock and the Herbig-Haro objects}. Basically
all data have been obtained with the IUE satellite (see
G\'omez de Castro \& Robles 2001 for a compilation).
Recent observations obtained with  Hopkins Ultraviolet Telescope
(HUT) have shown that it is still unclear how line
radiation is excited, at least, in HH2; in particular
O~VI is not detected as was expected in high
excitation Herbig-Haro objects where line radiation is
predicted to be produced in strong radiative shocks
where the shock kinetic energy is damped into heating
(Raymond et al 1997).

\end{enumerate}

\subsection{About the inner disk and ion belts}

Strong continuum FUV emission (1300--1700 \AA ) has
been detected recently from some stars with bright molecular disks
including GM~Aur, DM~Tau, and LkCa~15, together with inner disk
gaps of few AUs (Bergin et al. 2004). This emission is likely due
to energetic photoelectrons mixed into the molecular layer that
likely indicates the existence of a very hot component in the
inner disk.

High-resolution HST/STIS spectra have revealed, for the first
time, the rich UV molecular emission in CTTSs. H$_2$ fluorescence
emission has now been studied in detail in the nearest CTTS,
TW~Hya, and the richness of the spectrum is overwhelming: Herczeg
et al. (2002) detected 146 Lyman-band H$_2$ lines. The observed
emission is likely produced in the inner accretion disk, as are
the infrared CO and H$_2$O lines.   From these UV data, Herczeg et
al. (2004) estimated that the warm disk surface has a column
density of $N_{H_2} = 3.2 \times 10^{18}$~cm$^{-2}$, temperature
of $T=2500$~K, and filling factor of H$_2$ as seen from the source
of the Ly$\alpha$ emission of 0.25$\pm$0.08. The observed spectrum
shows that some ground electronic state H$_2$ levels with
excitation energies as large as 3.8 eV are pumped by Ly$\alpha$.
These highly excited levels may be formed by dissociative
recombination of H$^+_3$, which in turn may be formed by reactions
involving X-rays and UV photons from the star. Also DF~Tau and
V836~Tau H$_2$ emission seems to arise from the disk (Herczeg et
al 2006).

In addition to this molecular component, there is increasing
evidence of the existence of ion belts/rings around some TTSs. An
ion belt has been detected around the TTS, RW~Aur (G\'omez de
Castro \& Verdugo, 2003). A corotation radius of 4.4~$R_*$ is
derived and a $\log T_e (K) \simeq 4.7$ and $\log n_e (cm^{-3}) =
11.6$ are estimated. This was the first detection of such an
structure around a classical TTS. In addition, there are
indications of a similar structure around AB~Dor, a weak line TTS
(see Fig.~7). The structure is resolved, as in RW~Aur, because
there is an inner hole that allows separating the stellar/wind
contribution from the belt. However in a 5.7 hours time lapse the
double peaked profile is lost, and  the inner part of the profile
is filled in again with emission (G\'omez de Castro 2002) .

\begin{figure}
\centerline{\includegraphics[width=17pc]{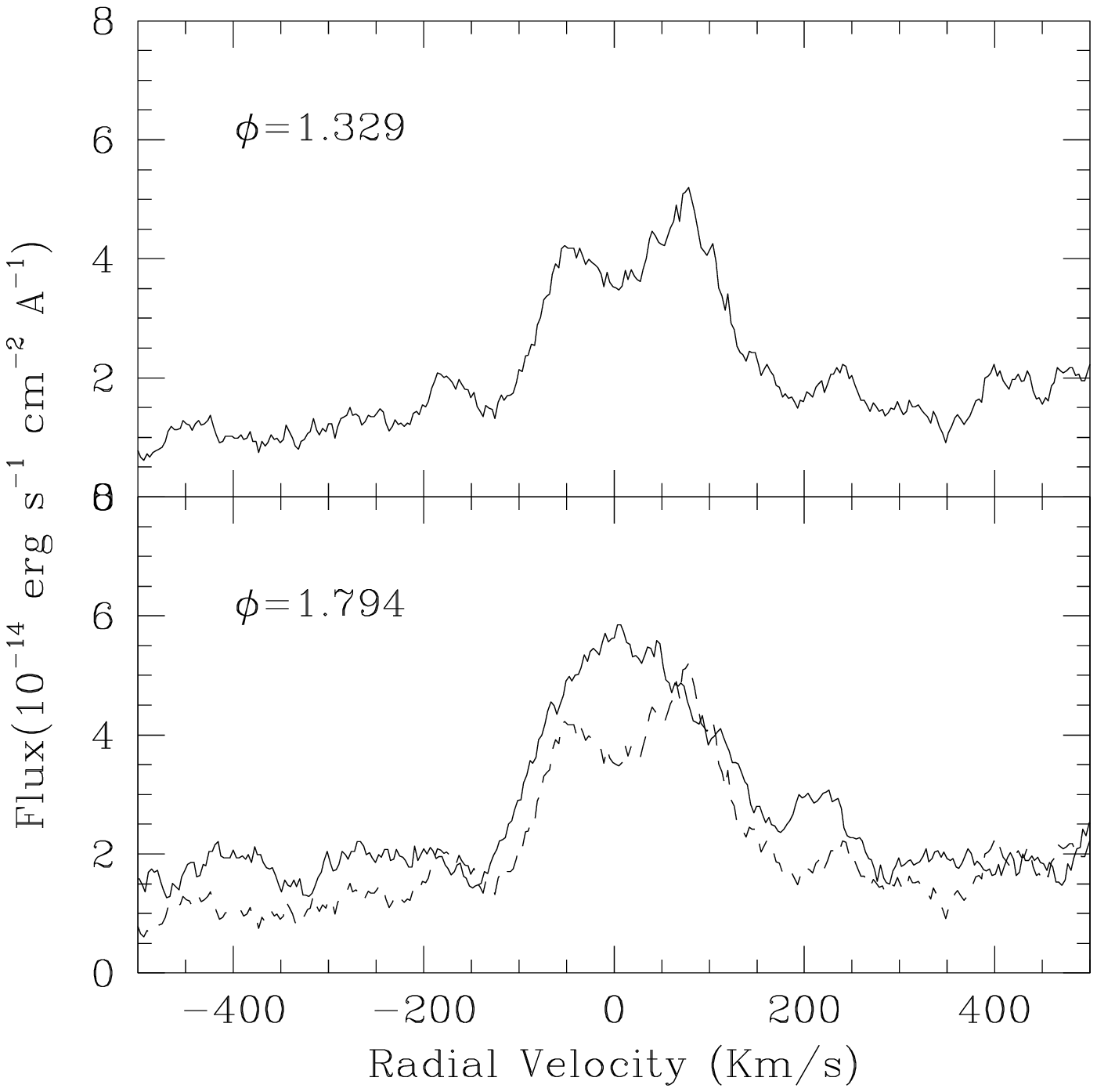}}
\caption[]{SiIII] profiles of AB~Dor obtained with the HST/GHRS
(see G\'omez de Castro 2002 for more details). In the bottom
panel, the profile at phase ($\phi$) 0.329 is overplotted
(dashed line) on the profiles at $\phi = 0.794$ (continuous line),
for comparison.}
\end{figure}

\subsection{About the interaction between the disk and the wind}

AB~Dor, a very bright nearby 30~Myr old star, is the only young
star that has been well monitored in the UV for flares. Nine events were
detected during 10.63~hours of monitoring with HST/GHRS!.
The C~IV and Si~IV UV line profiles produced by most of the events
are narrow and redshifted, indicating hot gas falling onto the
star during the flare. However, the strongest event produced a
very broad profile with narrow absorption slightly blueshifted.
This profile lasted a few kiloseconds and thus the broad wings are
most likely tracing the front shock of a CIR (G\'omez de Castro
2002). In the solar system, there are three very different types
of  ``flares'', which are sudden increases of the high energy
radiation and particles flux: magnetic flares (magnetic
reconnection events), corotating interaction regions or CIRs
(shock fronts formed by the interaction between the slow and the
fast component of the solar wind), and coronal mass ejections.
This classification also applies to TTSs and their circumstellar
environments. High-resolution UV spectroscopic monitoring is
required to disentangle the possible mechanisms for flares in
proto-stellar systems and to study their impact in young planetary disks
evolution as well as on planetary atmospheres embryos.

\begin{figure}
\centerline{\includegraphics[width=18pc]{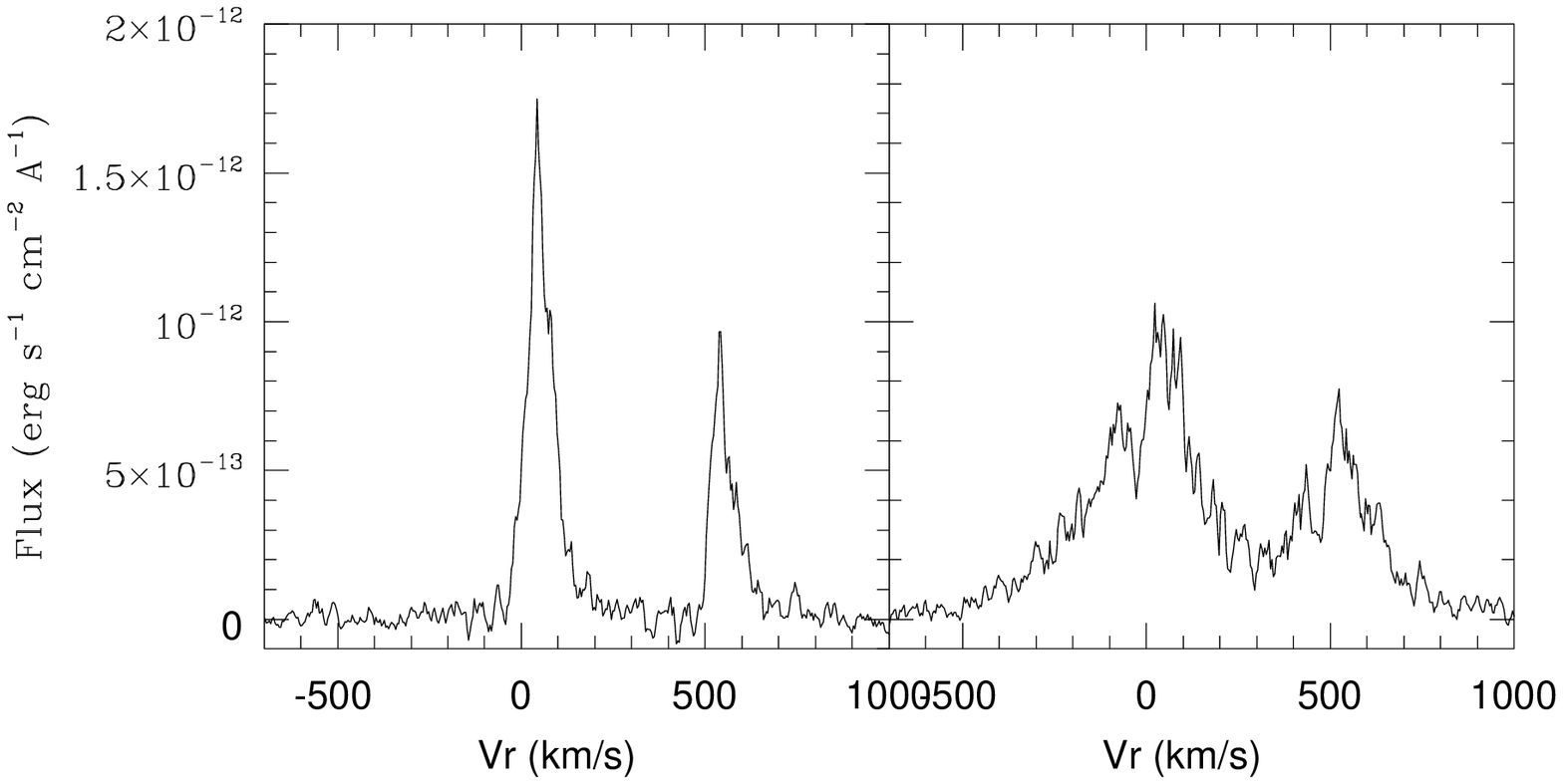}}
\caption[]{The C~IV 1548~\AA\ profile of AB~Dor during a normal
stellar flare (left) and a transient feature probably associated
with a CIR (right). Both events lasted several kiloseconds. The
left profile is typical of three events that occured during the
short monitoring time, while the profile on the right was observed
only once. Note the presence of a narrow absorption and the very
broad line wings in the right panel profile (see G\'omez de Castro
2002 for more details). }
\end{figure}

\section{Summary: the key observables}

In brief, the radiation produced by the accretion engine
(including magnetospheres, outflows, accretion, inner disk
and shock between winds and young planetary disks) is produced
in the UV. To separate the various contributions is necessary
either very high spatial resolution or moderate time resolution.

Current surveys show that although there are some few nearby TTSs
and WTTSs sparsely distributed around the Sun (AB~Dor at 14.9~pc
or TW~Hya at 56~pc) the nearest star forming complexes are
concentrated in a {\it Star Formation Belt} (SFB) at  140~pc
around the Sun which includes Taurus, Auriga-Perseus, Ophiuchus,
Lupus and Chamaleon molecular clouds and several thousands of
pre-main sequence stars forming in various environments (clustered
as in Ophichus, sparse as in Taurus). Resolving spatial scales of
a tenth of the solar radius in the SFB would allow to study the
connection between the star and the outflow in full detail. For
this purposes spatial resolutions of 3.3 micro arcseconds are
required; thus, for a fiducial wavelength of 1500\AA\ the aperture
must be about 10~km. Such a long baseline interferometry should be
carried out in the space and the Stellar Imager project (Carpenter
et al 2008) represents a first atempt to such an ambitious
project.

However, the requirements are not so strong to resolve the inner
disk structure, the disk wind and the plasmoids ejection from the
current layer between the stellar and the disk wind. In such a
case, spatial resolutions of 1-0.5 milliarcseconds (mas) would be
enough to map the SFB sources thus requiring apertures of 20~m and
effective areas about 10$^4$ times those of HST/STIS; research in
new coatings and detectors in the UV field (Kappelmann \&
Barnstedt, 2007) as well as a clever optical design may account
for a factor of ten but, still, a larger, $\sim 30$m aperture will
be required to get SNR$\simeq 10$ in the C~IV line in reasonable
exposure times (few hours). There is an ongoing project that
satisfies these requirements, the Fresnel Interferometer (see
Koechlin et al 2008) and some projections on the expected
performance of the interferometer on the mapping of the engine are
plotted in Fig.~9.

Both space interferometer projects are under study by their
national space agencies and, in case they succeed, they will
be available about 2030. Is there anything else to be done
{\it in the meantime}?. The answer is definitely positive, time
mapping will allow us to resolve the structures since the
variability time scales are not the same for all the
phenomena and they do not produce the same inprint in the
spectra (neither in temperatures, densities or velocities).
Some examples of the power of this technique have
already been shown in this contribution.

High resolution spectroscopy (R$\sim 50,000$) is enough to
discriminate among the various components thus, scaling with
the fluxes of the weakest H$_2$ lines (from Heczeg et al
2002) detected with the HST/STIS, a factor of 10 increase
of the effective area with respect to HST/STIS is required
to reach most of the sources
in the SFB. The COS instrument in HST will provide
a factor of 10 sensitivity increase with respect to STIS
in the 1150-1700\AA\ range because of its optimized optical
design (see Froening et al 2008), unfortunately
the orbital constrains of HST do not favour monitoring
programs. High orbit missions alike the WSO-UV (see Shustov et al
2008) are better suited for this purpose.

An additional factor of 10
would be required to obtain SNR$\sim 10$ in exposure times
of few minutes; this short time scales are necessary to
map variations in flare time scales as shown in
Fig.~8 for the pre-main sequence stars in the SFB.
Unfortunately, spectroscopic monitorings of the flaring activity in the
SFB will have to wait for future missions, with collecting
areas about 8-10~m, preferably located at the L2.

\begin{figure*}
\centerline{\includegraphics[width=26pc]{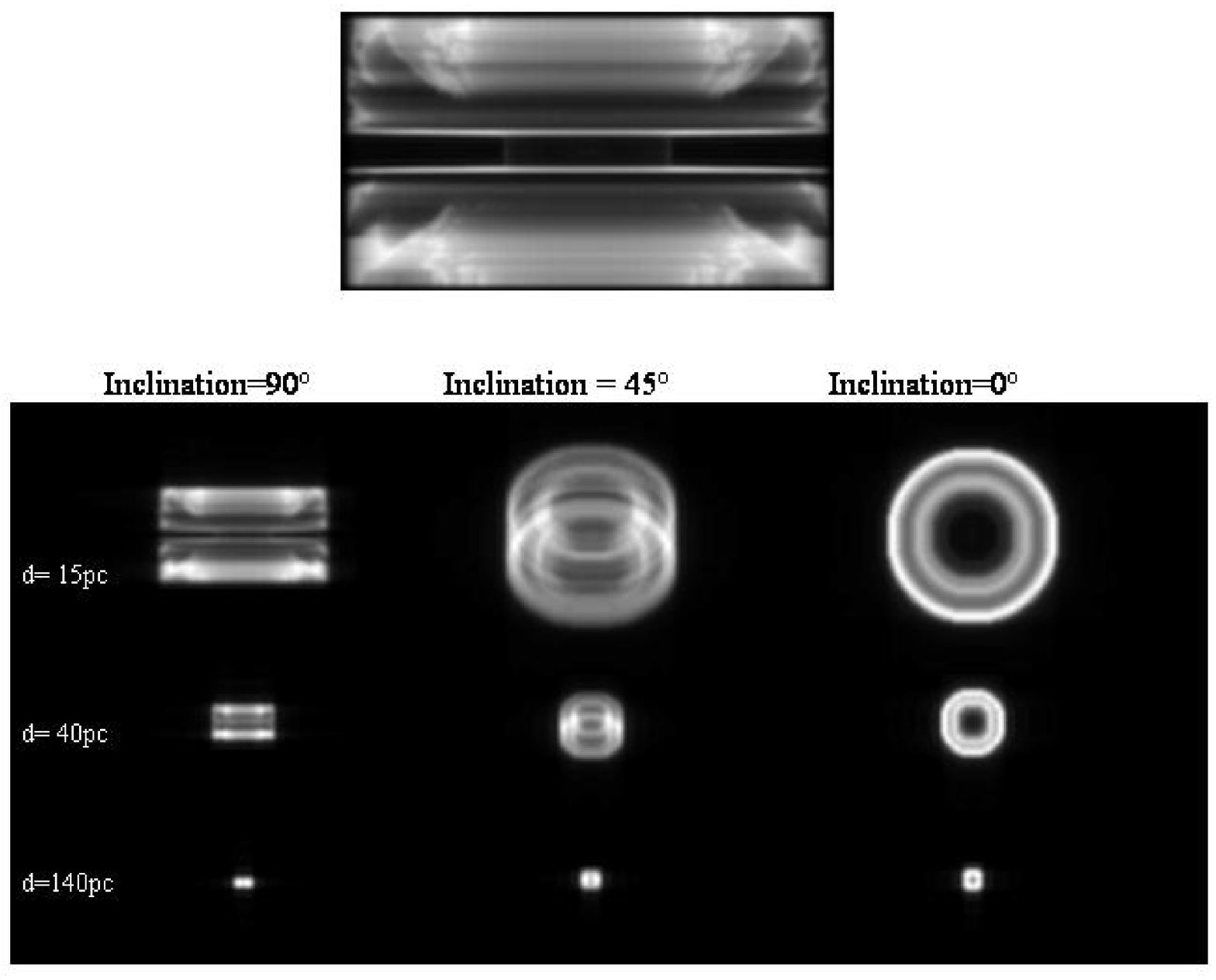}}
\caption[]{Theoretical prediction of the Si~III] emissivity from
numerical simulations of star-disk interaction (G\'omez de Castro
\& Rekowski, 2008). The stellar magnetosphere is assumed to be
dipolar with a field strength at the surface of 1kG. The
magnetosphere interacts with the disk which is under a moderate
$\alpha$-dynamo effect\footnote{The magnetic field in the disk is
assumed to be generated by a standard $\alpha ^2 \Omega $ dynamo
(e.g. Krauser \& Raedler 1980) where $\alpha$ is the mean-field
$\alpha$ effect and $\Omega$ the angular velocity of the plasma.
$\alpha$ scaling is: $\alpha = -0.1 \frac{z}{z_0} \frac {\chi
_{\rm disk} (r, z)}{1+ V_A^2/C_s^2}$ where $\chi _{\rm disk}$ is
the disk profile (see von Rekowski et al 2003), $z_0$ is the disk
half-thickness, $V_A$ is the local Alfv\'en velocity and $C_s$ is
the local sound speed.} . The inner disk wind is
magnetocentrifugally accelerated ({\it top panel}).

The convolution of the theoretical prediction with the point
spread function of the 30m Fresnel Interferometer
FII has been carried out by Laurent Koechlin
and Truswin Raksasataya.
The convolution is shown for three inclinations
(0$^o$, 45$^o$ and 90$^o$) and three distances to the Earth
(15~pc, 40~pc and 140~pc). Notice that the inner ring is resolved
even for 140~pc.
 }
\end{figure*}

%
\acknowledgments

This work has been supported by the Ministry of Education of Spain
through grant AYA2007-67726 and the Comunidad Aut\'onoma de Madrid
through grant CAM-S-0505/ESP/0237


%

%

\end{document}